%% file: xcache-icnc23.tex
\documentclass[conference]{IEEEtran}
\IEEEoverridecommandlockouts

\usepackage{cite}
\usepackage{amsmath,amsfonts}
\usepackage{algorithmic}
\usepackage{graphicx}
\usepackage[caption=false]{subfig}
\usepackage{float}
\usepackage{multirow}
\usepackage{array}
\usepackage{textcomp}
\usepackage[hyperindex,breaklinks,hidelinks]{hyperref}
\hypersetup{%
   pdftitle={Effectiveness and predictability of in-network storage cache for Scientific Workflows},%
   pdfauthor={Caitlin Sim, Kesheng Wu, Alex Sim, Inder Monga, Chin Guok, Frank Wurthwein, Diego Davila, Harvey Newman, Justas Balcas},
   pdfkeywords={network utilization, storage cache, data service,
     predictability, AI for networking},
   colorlinks=true,     
   linkcolor=blue,      
   citecolor=blue,      
   filecolor=blue,      
   urlcolor=blue        
}

\graphicspath{{figs}}

\usepackage{xcolor}
\def\BibTeX{{\rm B\kern-.05em{\sc i\kern-.025em b}\kern-.08em
    T\kern-.1667em\lower.7ex\hbox{E}\kern-.125emX}}

\begin{document}

\title{Effectiveness and predictability of in-network storage cache for Scientific Workflows}

\author{
\IEEEauthorblockN{Caitlin Sim$^1$, Kesheng Wu$^2$, Alex Sim$^2$, Inder Monga$^3$, Chin Guok$^3$\\Frank W\"{u}rthwein$^4$, Diego Davila$^4$, Harvey Newman$^5$, Justas Balcas$^5$}
\IEEEauthorblockA{\\$^1$ University of California at Berkeley, caitlinsim@berkeley.edu\\
$^2$ Lawrence Berkeley National Laboratory, \{kwu,asim\}@lbl.gov\\
$^3$ Energy Sciences Network, \{imonga,chin\}@es.net\\
$^4$ University of California at San Diego, \{fkw,didavila\}@ucsd.edu\\
$^5$ California Institute of Technology, newman@hep.caltech.edu, jbalcas@caltech.edu
}
}

\maketitle

\begin{abstract}
Large scientific collaborations often have multiple scientists accessing
the same set of files while doing different analyses,
which create repeated accesses to the large amounts of shared data
located far away.
These data accesses have long latency due to distance and occupy the limited bandwidth available over the wide-area network.
To reduce the wide-area network traffic and the data access latency,
regional data storage caches have been installed as a new networking
service.
To study the effectiveness of such a cache system in scientific
applications, we examine the Southern California Petabyte Scale Cache for
a high-energy physics experiment.
By examining about 3TB of operational logs, we show
that this cache removed 67.6\% of file requests from the wide-area
network and reduced the traffic volume on wide-area network 
by 12.3TB (or 35.4\%) an average day.
The reduction in the traffic volume (35.4\%) is less than the reduction in file
counts (67.6\%) because the larger files are less likely to be reused.
Due to this difference in data access patterns, 
the cache system has implemented a policy
to avoid evicting smaller files when processing larger files.
We also build a machine learning model to study the predictability of
the cache behavior.
Tests show that this model is able to accurately predict the cache accesses, cache
misses, and network throughput, making the model
useful for future studies on resource provisioning and planning.
\end{abstract}

\begin{IEEEkeywords}
in-network caching, data throughput, transfer performance, data access trends
\end{IEEEkeywords}

\section{Introduction}
\label{sec:intro}
\input{intro}

\section{Background and XCache Log Files}
\label{sec:data}
\input{data}

\section{Network Traffic Reduction}
\label{sec:stat}
\input{stat}

\section{Modeling Transfer Throughput}
\label{sec:lstm}
\input{lstm}

\section{Conclusion}
\label{sec:conc}
In this study, we set out to understand the effectiveness of in-network
storage cache used by a distributed scientific collaboration.
The source information is a set of 3TB of operational logs from the XCache
servers on SoCal Repo.
The data analysis operations of the collaboration commonly involve two
types of files, the smaller sized ones are used frequently with more reuse, while the larger sized 
ones are invoked infrequently with less reuse.
We observed that SoCal Repo could on average serve about 67.6\% of
files from its disk cache, while on average only 35.4\% of bytes
requested could be served from the cache, because the large files are
less likely to be reused.
To avoid cache pollution from this usage pattern of large files, the
system operators have separated the two different types of files with different storage nodes.
During the period where fewer large files are requested (3/2022 -- 5/2022), the wide-area
network traffic is reduced by about 29TB per day.
Over the whole period of observation, there is a five-month period where
the large file requests are noticeably high.
The average reduction of wide-area network traffic by this cache over
the whole observation period is still about 12.3TB per day, which is quite
significant.

This work also explored an option to model the network performances with
a neural network architecture known as LSTM.
Tests show that the prediction error (measured as RMSE) are quite small.
In a case where the original time series has large variations, we also
show that the LSTM model could work quite well on moving-averaged
versions of the time series data.
With this model, we plan to consider how to provision future deployments
of in-network caches.
We are also planning to study other storage caches currently under
deployment to gain better understanding of in-network caches.

\section*{Acknowledgment}
This work was supported by the Office of Advanced Scientific Computing Research, Office of Science, of the US Department of Energy under Contract No. DE-AC02-05CH11231, and also used resources of the National Energy Research Scientific Computing Center (NERSC). 
This work was also supported by the National Science Foundation through the grants OAC-2030508, OAC-1836650, MPS-1148698, PHY-2121686 and OAC-1541349.
This work was also supported in part by the US Department of Energy, Office of Science, Office of Workforce Development for Teachers and Scientists (WDTS) under the Science Undergraduate Laboratory Internship (SULI) program.

\bibliographystyle{IEEEtran}
\bibliography{papers}

\end{document}

%% file: intro.tex
Large scientific projects often involve thousands of scientists
sharing a massive data collection~\cite{esnetHepReq}.
These projects, such as the Large Hadron Collider (LHC), have
collaborators around the world, each with their own analysis tasks,
accessing a different portion of the data collection, transferring data files over
long distances, and causing high demand over the limited wide-area network bandwidth.
While efficient tools for data movement over wide-area network are
available, there is a new networking service, in-network storage caches,
that could reduce a significant portion of the traffic on wide-area network.
These caches take
advantage of the geographical sharing of data accesses as there are
overlapping in data accesses among the colleagues in the
same institution who often work on related scientific objectives.
In particular, the High Energy Physics (HEP) community has been exploring such a
caching system under the term of regional ``data lakes''~\cite{datalakes}
as a part of their federated data storage infrastructure~\cite{xrootdcms}.
There are evidences that a regional data cache could improve data accesses~\cite{Fajardo2020, Kosar:2011:stork, Tierney:1999:cache}. 
However, real-world deployment sometimes bring up unexpected challenges.
This work studies the effectiveness of one such in-network deployment to see how to address the challenges encountered.

This work studies the operational logs of a large-scale deployment of
storage cache nodes. 
These logs are from the Southern California
Petabyte Scale Cache (SoCal Repo)~\cite{socalrepo2018} developed for
High-Energy Physics (HEP) data analysis jobs,
where the wide-area network traffic is primarily carried by the Energy Sciences Network (ESnet)~\cite{esnet2022}.
There have been some reports about the performance characteristics
including number of file requests, cache misses, and data
volumes~\cite{copps2021, han2022}.
The first objective of this work is to understand the networking
characteristics such as network traffic reduction, data
throughput performance, and so on.
We expect this part of the study to confirm that SoCal Repo
significantly reduce the traffic over the wide-area network.
Nevertheless, there are surprises due to a special user access pattern.

The second objective of our work is to understand the predictability of
the network utilization to help plan for additional
deployments of in-network caches in the science network infrastructure.
For this purpose, we developed a machine learning model to predict the network utilization.
Despite the high variability in the cache usage, as shown in
Figure~\ref{fig_1}, it is still possible to model the cache requests with accuracy.
Our model takes the SoCal Repo performance characteristics as the input
time series and learns the performance patterns through a recurrent
neural network architecture known as the Long Short-Term Memory
(LSTM)~\cite{greff2016lstm, sherstinsky2020fundamentals}.
The errors of the predictions are significantly less than the standard deviation of the original values.
One way to use this prediction model is to plan for days with unusually high network demands and maximize the overall system performance.

%% file: data.tex
SoCal Repo is a storage cache supporting computing jobs in Southern
California for US Compact Muon Solenoid (CMS) experiment, a HEP
collaboration with participants around the world~\cite{xrootd2005, xrootdcms}.
The analysis jobs involve files of different types, for example,
analysis object data (AOD), MiniAOD, or NanoAOD files, where 
the information content per proton collision event differs by more than a factor of 10 each going from AOD to MiniAOD to NanoAOD.
NanoAOD is thus O(10,000) smaller than AOD.
More than 90\% of analyses work with either MiniAOD or NanoAOD~\cite{miniaod2019, cmsoffline}.
The analysis work mostly starts with exploration of NanoAOD or MiniAOD files.
In a number of cases, after scientists have determined the most
useful algorithms and found the most promising collision events for
their analysis work, they apply some algorithms on the larger data
file formats (which has more detailed information about the selected events) to produce the final results.
This data usage pattern effectively creates two types of accesses, one
type touches small file formats frequently and the other retrieves large
infrequently.
The analysis jobs requiring large file formats might take a considerable amount of time because the file transfers, including potentially retrieval from tape archives, and computation both are time consuming.

The SoCal Repo
has approximately 2.5PB of total storage with 24 federated
caching nodes.
There are 11 nodes at Caltech with storage sizes ranging from 96TB to
388TB, 12 nodes at UCSD with 24TB each node, and one node at ESnet
Sunnyvale endpoint with 44TB of storage.
Furthest distance to the cache node from the computing resources is
about 500 miles from UCSD to ESnet Sunnyvale endpoint, with an RTT of
about 10ms.
The measurement data has been collected from July 2021 through June 2022,
consisting of 8.7 million data accesses where 67.6\% are satisfied with files in cache, see \autoref{tb:summary} for
additional summary statistics.
Among the 12.7 PB requested, 4.5 PB (35.4\%) could be served from the cache, while 8.2 PB needs to be transferred over wide-area network.
The difference between 67.6\% and 35.4\% is one of the issues we seek to resolve.

\begin{table}[htb!]
\scriptsize
\centering
\caption{Summary of data access from July 2021 to June 2022.  About
  67.6\% of file requests are satisfied by this cache, while 35.4\%
  requested bytes are in the cache.}
\begin{tabular}{|l||r|r|r|r|r|} \hline
  & {\shortstack{\# of\\accesses}} & {\shortstack{cache miss\\size (TB)}} & {\shortstack{cache hit\\size (TB)}} & {\shortstack{number of\\cache misses}} & {\shortstack{number of\\cache hits}} \tabularnewline \hline \hline
Total & 8,713,894 & 8,210.78 & 4,499.44 & 2,822,014 & 5,891,880 \tabularnewline \hline
Daily & 23,808 & 22.43 & 12.29 & 7,710 & 16,098 \tabularnewline \hline
\end{tabular}
\label{tb:summary}
\end{table}

Cache misses occur when the client's requested data file is not in any
of the cache nodes and needs to be transferred from a remote storage
over the wide-area network.  When the client's requested data is in one
of the cache nodes, it is a cache hit, and the data is served from the
cache without a wide-area data transfer.
The network traffic reduction comes from these cache hits.

The cache nodes run on XCache software~\cite{stashcache, xcache2014, socalrepo2018}.
The information used in this study is extracted from XCache log files.  We've
processed 8,433 log files amounting to about 3 TB, and
extracted information about the request sizes, how the request is
satisfied, etc.
From such information, we derived cache hits, cache misses, along with
network performance information such as remote transfer throughput that are used in later sections.

%% file: stat.tex
\begin{figure}
\centerline{\subfloat[Fraction of daily file requests: cache misses (in \textcolor{orange}{orange}) and
cache hits (in \textcolor{blue}{blue})]{
  \includegraphics[width=0.8\linewidth]{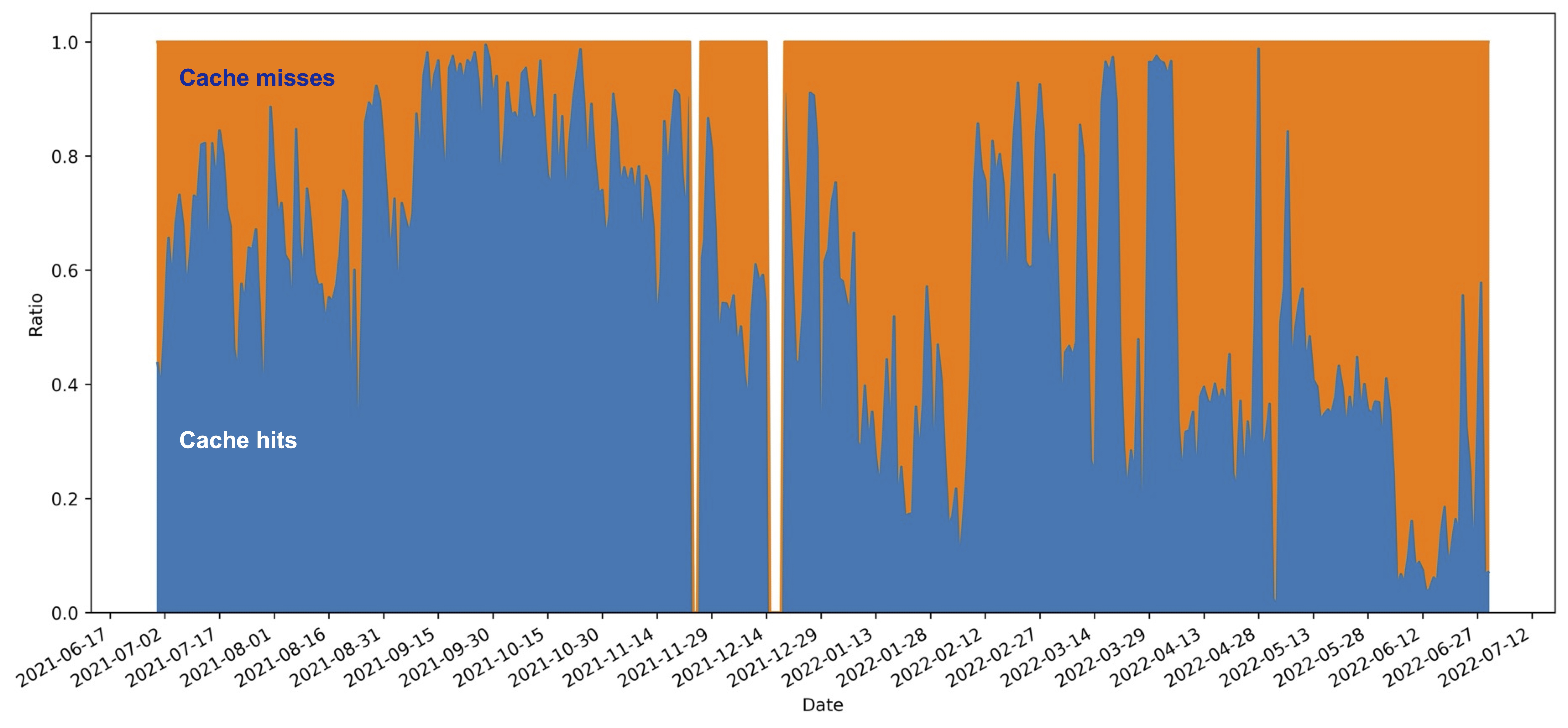}
  \label{fig:1a}}}
\centerline{\subfloat[Fraction of daily requested bytes: cache misses (in \textcolor{orange}{orange}) and
cache hits (in \textcolor{blue}{blue})]{
  \includegraphics[width=0.8\linewidth]{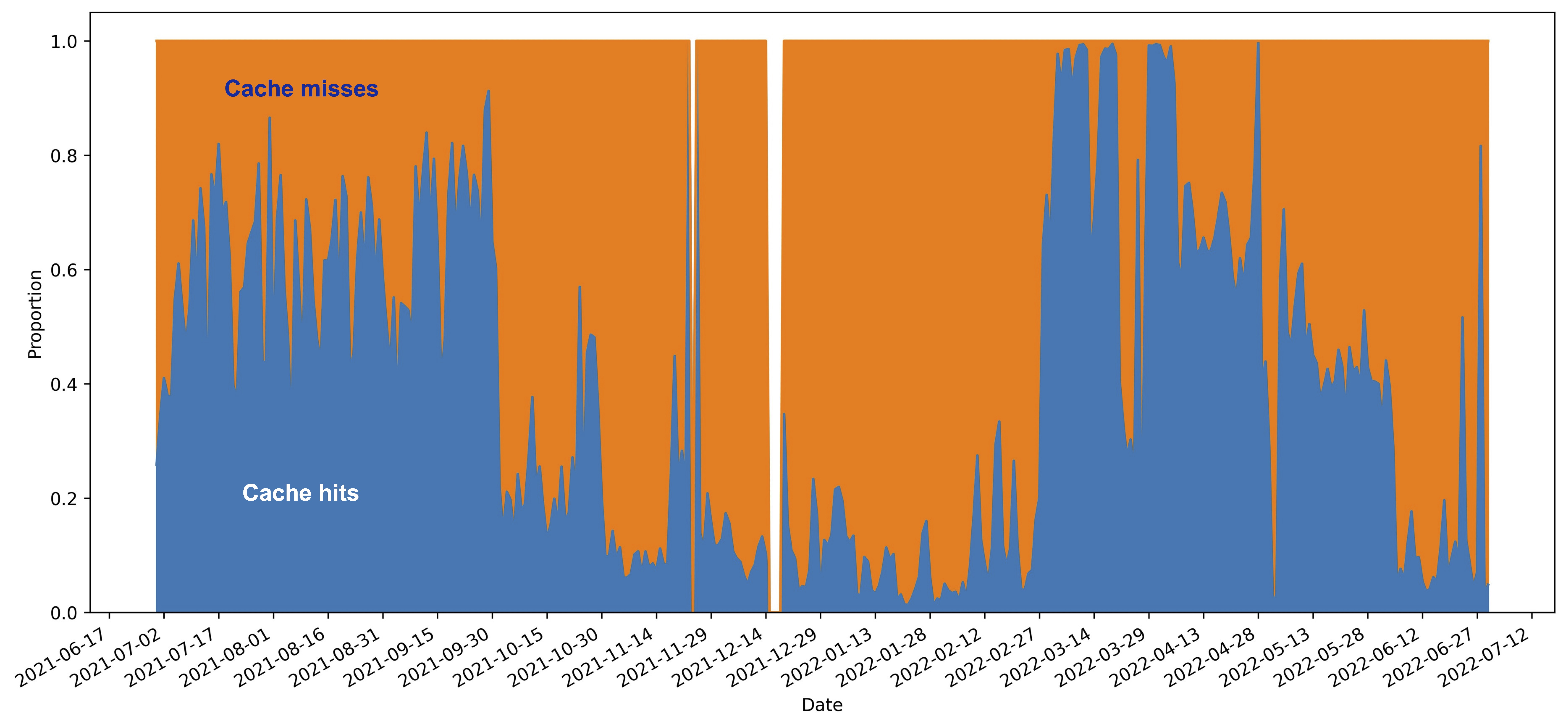}
  \label{fig:1b}}}
\caption{Cache miss rates versus cache hit rates based on (a) files
  requested and (b) bytes requested.}
\label{fig_1}
\end{figure}

\begin{figure}
\centerline{%
\subfloat[Daily file requests (count): cache misses (in \textcolor{orange}{orange}) and
cache hits (in \textcolor{blue}{blue})]{
    \includegraphics[width=0.8\linewidth]{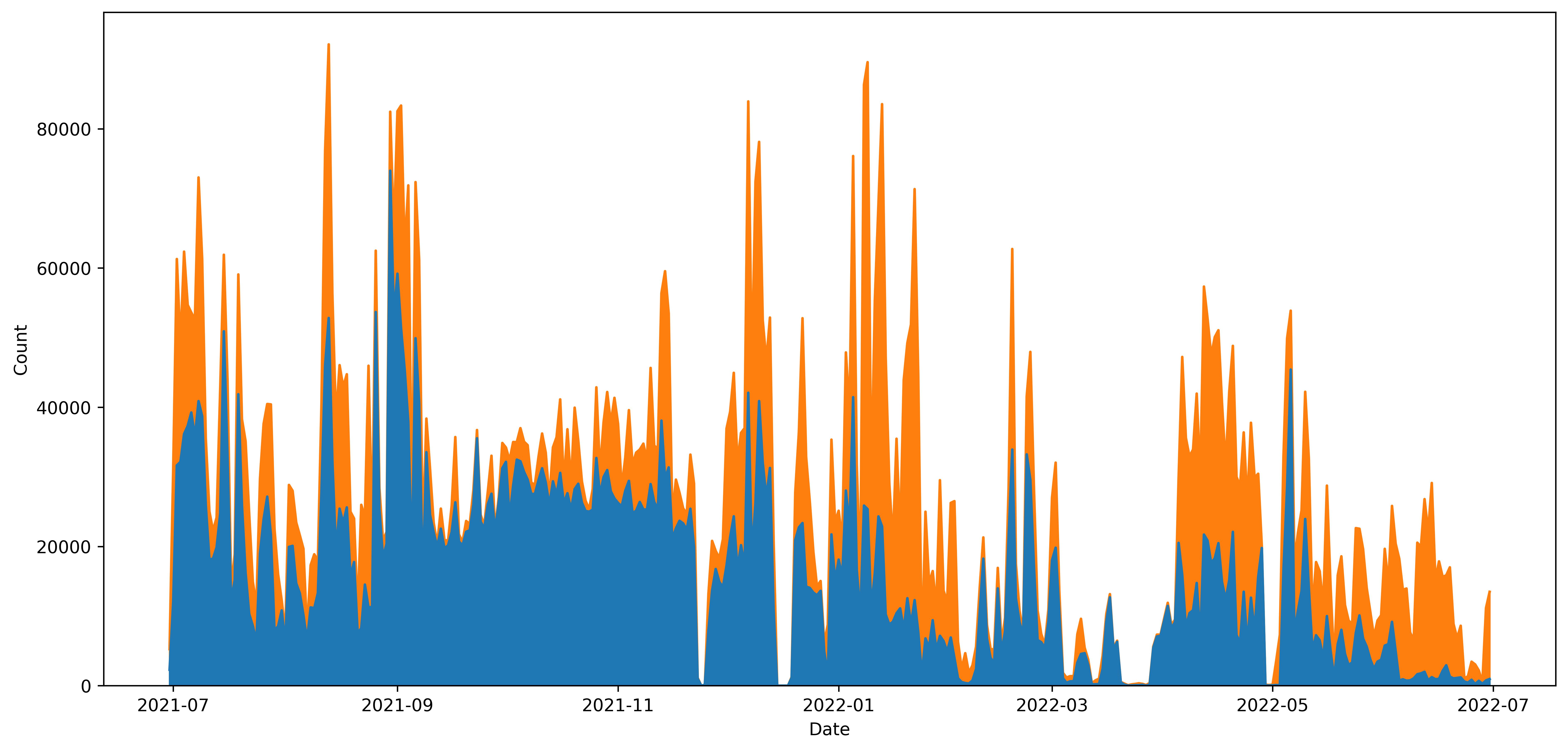}
    \label{fig_2a}
}}
\centerline{%
\subfloat[Daily traffic volume: cache misses (in \textcolor{orange}{orange}) and
cache hits (in \textcolor{blue}{blue})]{%
    \includegraphics[width=0.8\linewidth]{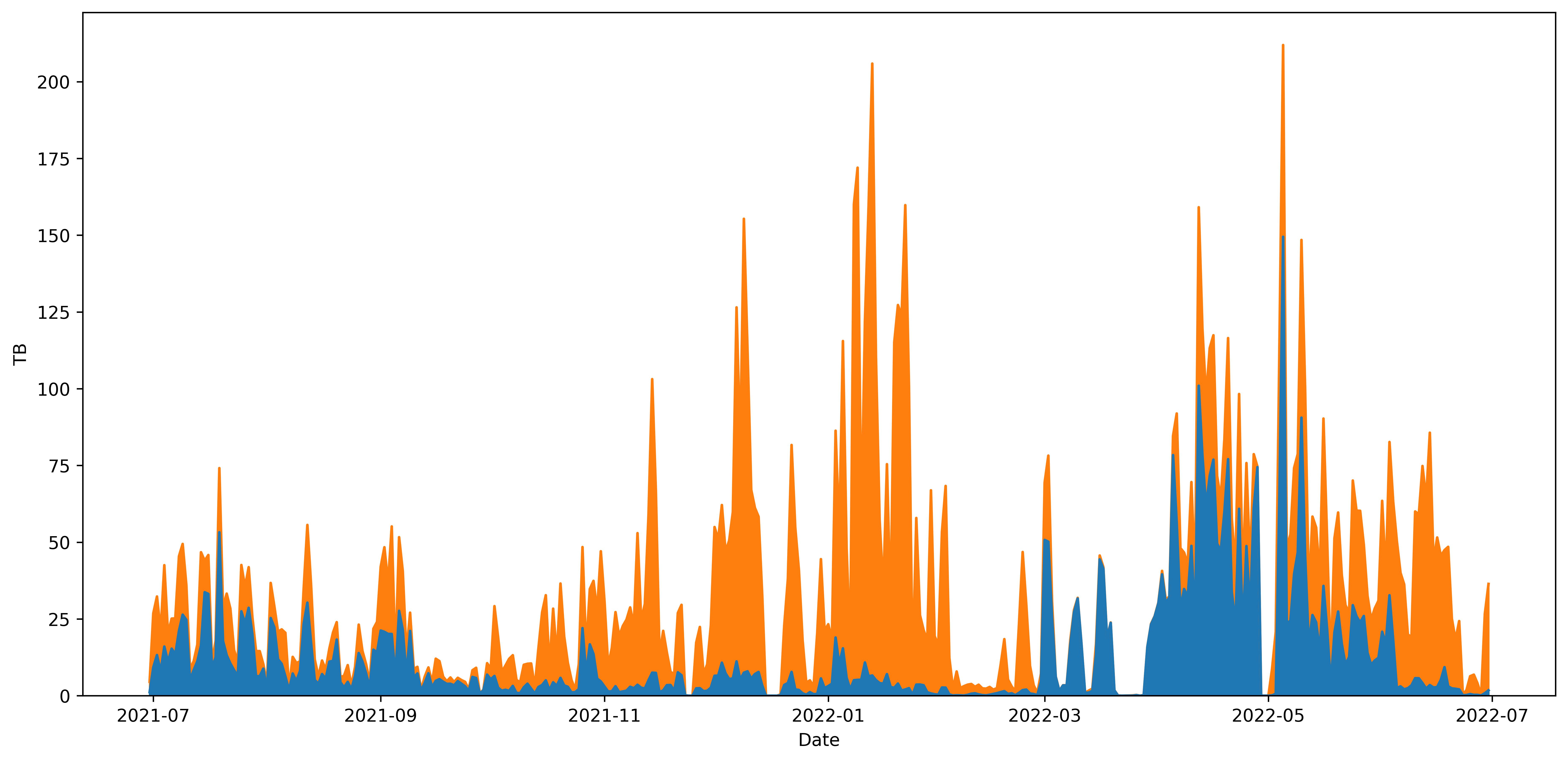}
    \label{fig_2b}
}} 
\caption{Daily network traffic statistics: (a) files requested and (b)
  bytes requested.  Note that only file requests
that miss the cache trigger remote network file transfers.}
\label{fig_2}
\end{figure}

In this section, we show
how much wide-area network traffic is actually saved by SoCal Repo.
\autoref{fig_1} shows the daily cache hit rate and cache miss rate,
where Figure~\ref{fig:1a} shows these rates based on the file request
counts and Figure~\ref{fig:1b} based on bytes requested.\footnote{There are
  two narrow gaps in \autoref{fig_1} due to brief periods of down time.}
Cache misses trigger wide-area network traffic. 
We show the cache miss rates with bright orange in \autoref{fig_1} and \autoref{fig_2}

Overall, the cache miss rates based on files requested (in
Figure~\ref{fig:1a}) are more stable than those based on bytes requested in Figure~\ref{fig:1b}.
In particular, there is a 5-month long period between Oct.~2021 and
Feb.~2022, where the majority of bytes requested are cache misses.
An examination of the number of files requested and number of bytes
requested in \autoref{fig_2} provides more information.


Figure~\ref{fig_2a} shows the number of files requested, separated into
those could be satisfied with files in the cache (hits, in blue) and those
require wide-area network transfers (misses, in orange).
Across all 24 cache nodes in the SoCal Repo, an average day sees about 16,000
file requests as hits along with 8,000 misses.
In terms of bytes requested, Figure~\ref{fig_2b} shows that about 12.3 TB per
day are served out of the cache during the whole year.
In the early part of the year, between Jul.~2021 and Sep.~2021, the
wide-area network traffic is reduced by about 13 TB per day, and between
March 2022 and May 2022, the wide-area network traffic is reduced by
about 29TB per day.

In the middle of our observation period, for example January 13, 2022, there are about 60,000 cache
misses amounting to about 200TB of wide-area traffic.  On average, each of
these files is over 3.3GB, which means they are large among
CMS data files.
This observation from the cache statistics conforms to the usage
patterns involving large files described in Section~\ref{sec:data}.
We also received additional confirmation from the site operators that these are indeed a
small number of data analyses involving large files.

This particular usage pattern involving large files has the potential of
evicting the smaller files (that are used more frequently)\footnote{This is colloquially known as cache pollution.} and reducing the overall effectiveness of the cache system.  
The operators of SoCal Repo recognized this usage pattern and have separated and limited the accesses to the 
cache nodes based on file types, which effectively prevents cache pollution.
In cases where one couldn't differentiate the cache usages based
on simple known characteristics, an alternative strategy
could be to have these requests
bypass the cache system~\cite{Malik:2005:bypass}.

%% file: lstm.tex
Now that we know the storage cache is effective in reducing the traffic
on wide-area network, and there are strategies to mitigate the impact of special access patterns that pollute the cache, we'd like to see how to provision additional storage
cache nodes in the future.  For this purpose, we start to model the
current cache usage and network performance.
More specifically, we build machine learning models for the hourly and
daily average data throughput performance 
as well as statistics about cache misses.
The data throughput is defined as the data transfer size over the
transfer time.
This information is useful for anticipating the time needed for file transfers.

We have decided to use a neural network architecture called
LSTM~\cite{greff2016lstm, sherstinsky2020fundamentals} because it is
effective in capturing time series patterns.
Our model includes the following features: cache miss count, cache miss
size, cache hit count, cache hit size, aggregate throughput for cache
misses on all nodes, aggregate throughput on cache hits for all nodes,
average throughput for each cache miss, and average throughput for each
cache hit.
In the remaining of this section, we discuss the information relates to cache misses, which are more relevant to the wide-area network performance.
The training data comes from the first 80\% of the whole monitoring
period, and the testing data comes from the last 20\%.
\autoref{LSTM_parameters} shows hyper-parameters chosen after exploring
about 1400 different parameter combinations.

\begin{table}
\centering
\caption{hyper-parameters of the LSTM models}
\begin{tabular}{|c|c|c|c|} \hline
{\shortstack{\# of\\LSTM unit}} & {\shortstack{activation\\function}} & {\shortstack{dropout\\rate}} & {\shortstack{\# of\\epochs}} \tabularnewline \hline \hline
128 & tanh & 0.04 & 50\tabularnewline \hline
\end{tabular}
\label{LSTM_parameters}
\end{table}

\begin{table}
\centering
\caption{RMSE of daily/hourly LSTM model results for network storage
  cache performance.  The relative prediction errors (inside
  parentheses) are measured against the standard deviations.  Note that
  all six rows are about cache misses.}
\scriptsize
\begin{tabular}{|c||r|r|r|} \hline
  & \shortstack{Training \\RMSE} & \shortstack{Testing \\RMSE} & \shortstack{standard \\deviation} \tabularnewline \hline \hline
Daily cache misses & 4306.01 & 3637.39 (.32) & 11317.08  \tabularnewline \hline
Hourly cache misses &  175.11 &   99.31 (.17) &   595.81  \tabularnewline \hline
\shortstack{Daily volume}  & 9.75 & 14.54 (.49) & 29.46  \tabularnewline \hline
\shortstack{Hourly volume} & 0.19 & 0.49 (.35) & 1.42  \tabularnewline \hline
\shortstack{Daily average throughput} & 33.20 & 23.49 (.21) & 110.43 \tabularnewline \hline
\shortstack{Hourly average throughput} & 27.08 & 22.79 (.19) & 121.36 \tabularnewline \hline
\end{tabular}
\label{tab:rmse}
\end{table}

As an overall performance measure, \autoref{tab:rmse} shows the
root-mean-square error (RMSE) of both the daily and hourly models for
the data volume and average (wide-area) network transfer performance.
The column labeled ``standard deviation'' is the standard deviation of the input
data values.  It provides a reference for us to determine how large are the
errors of predictions.  The ratios of testing RMSE and standard
deviation are shown inside parentheses.
In all cases shown, this relative error is less than 0.5, indicating the
predictions are pretty accurate.

In all three sets of measures shown in \autoref{tab:rmse}, the
LSTM models are more accurate with the hourly time series than with the
daily time series as both absolute and relative error are smaller.
The most likely reason might be there are more training data records
for the hourly time series.
Next, we look into more details of these prediction models.

\begin{figure}
\centerline{%
\subfloat[Daily number of cache misses]{
    \includegraphics[width=0.8\linewidth]{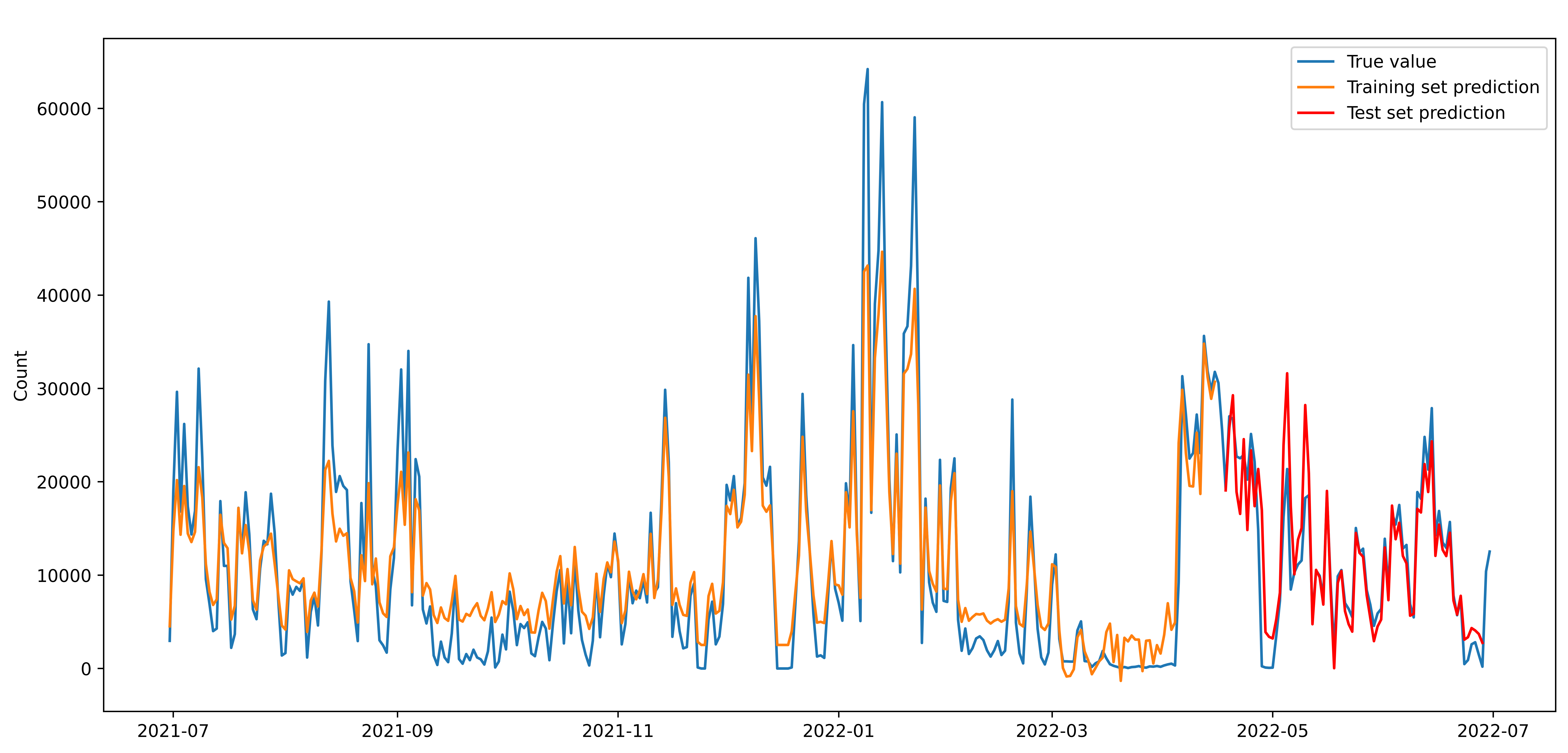}
    \label{fig:daily_miss_counts}
}}
\centerline{%
\subfloat[Hourly number of cache misses]{%
    \includegraphics[width=0.8\linewidth]{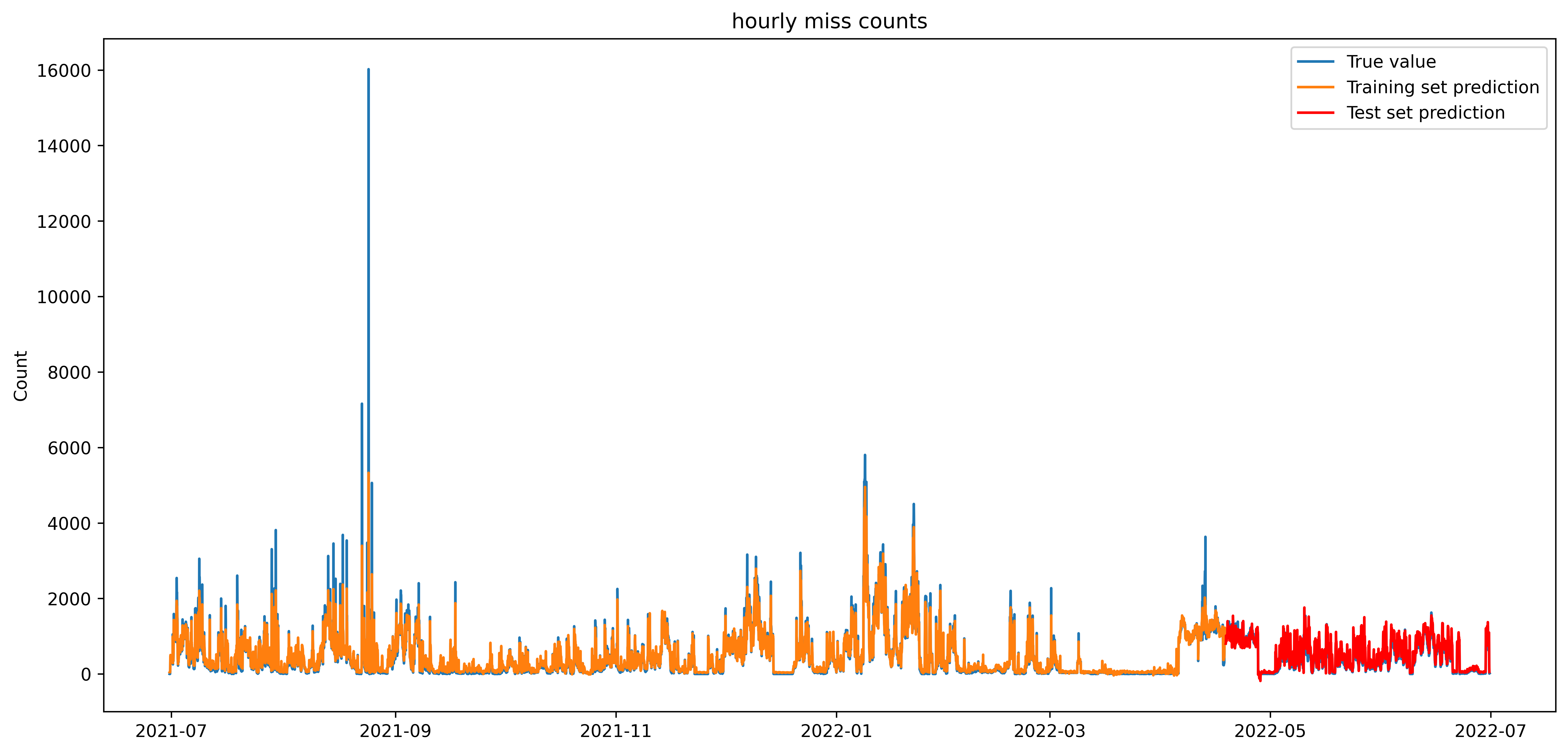}
    \label{fig:hourly_miss_counts}
}} 
\caption{Number of cache misses}
\label{fig_miss_counts}
\end{figure}

\begin{figure}
\centerline{%
\subfloat[Daily volume of cache misses]{
    \includegraphics[width=0.8\linewidth]{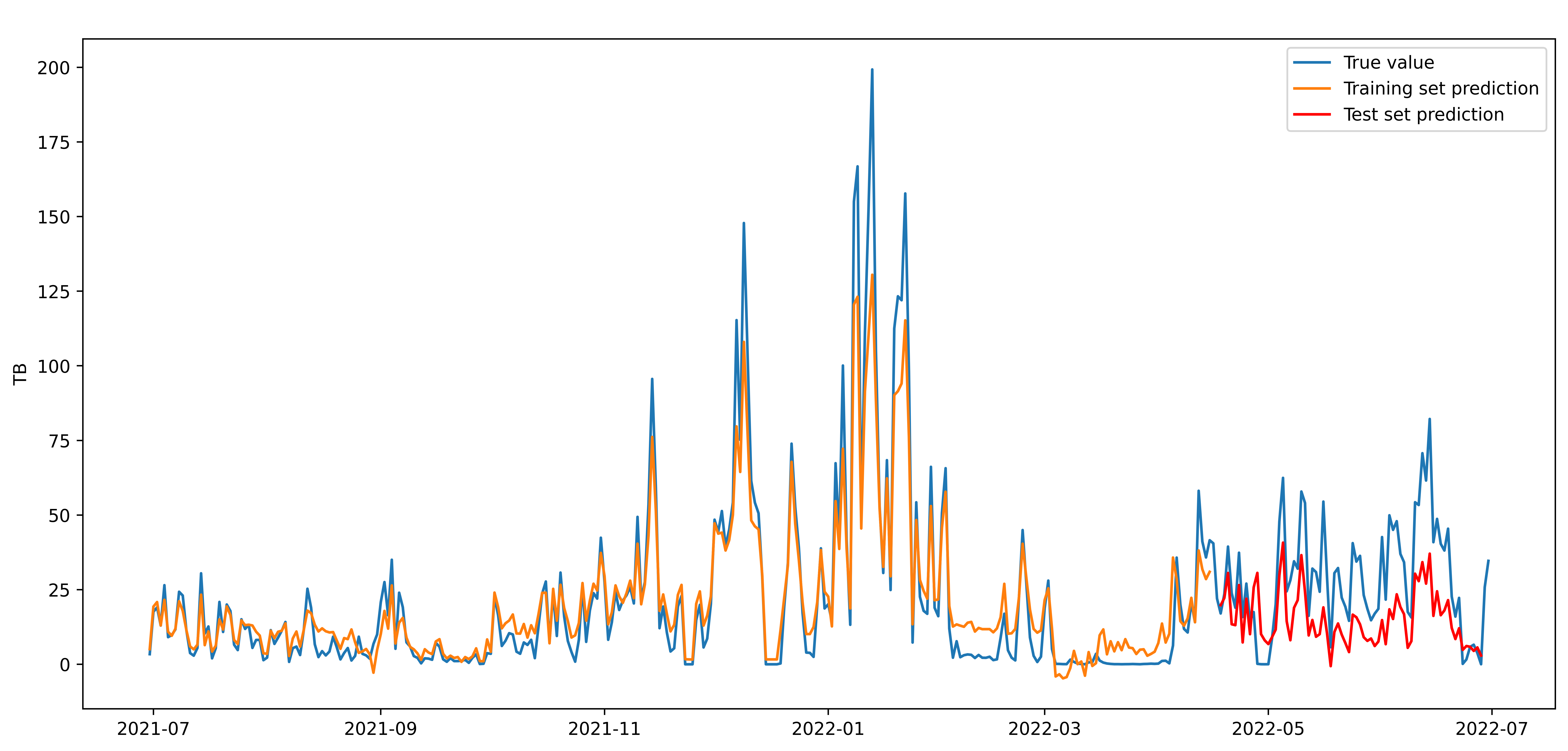}
    \label{fig:daily_miss_size}
}}
\centerline{%
\subfloat[Hourly volume of cache misses]{%
    \includegraphics[width=0.8\linewidth]{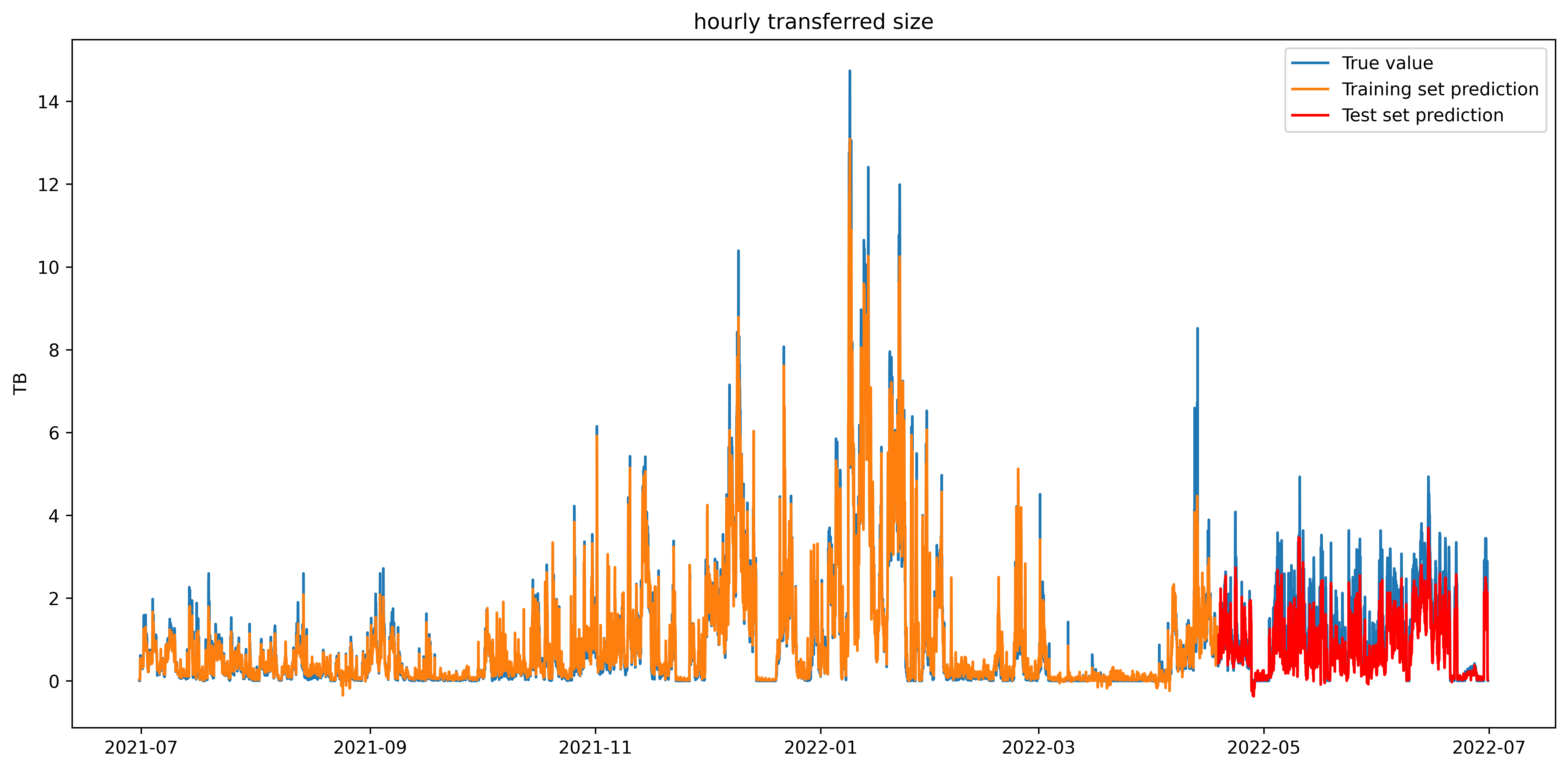}
    \label{fig:hourly_miss_size}
}} 
\caption{Bytes in cache misses}
\label{fig_miss_size}
\end{figure}

\begin{figure}
\centerline{%
\subfloat[Daily throughput of wide-area transfers]{
    \includegraphics[width=0.8\linewidth]{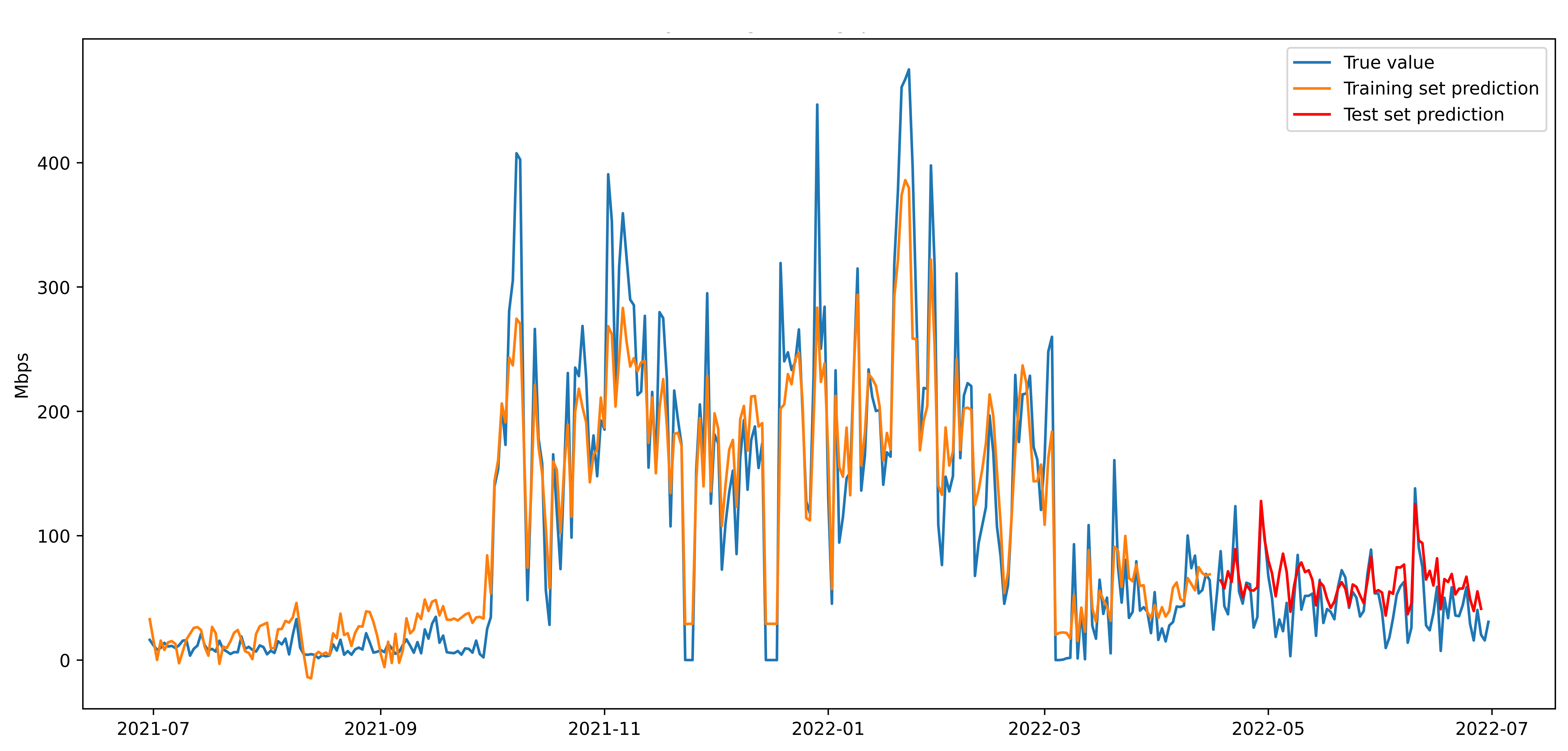}
    \label{fig:daily_miss_tput}
}}
\centerline{%
\subfloat[Hourly throughput of wide-area transfers]{%
    \includegraphics[width=0.8\linewidth]{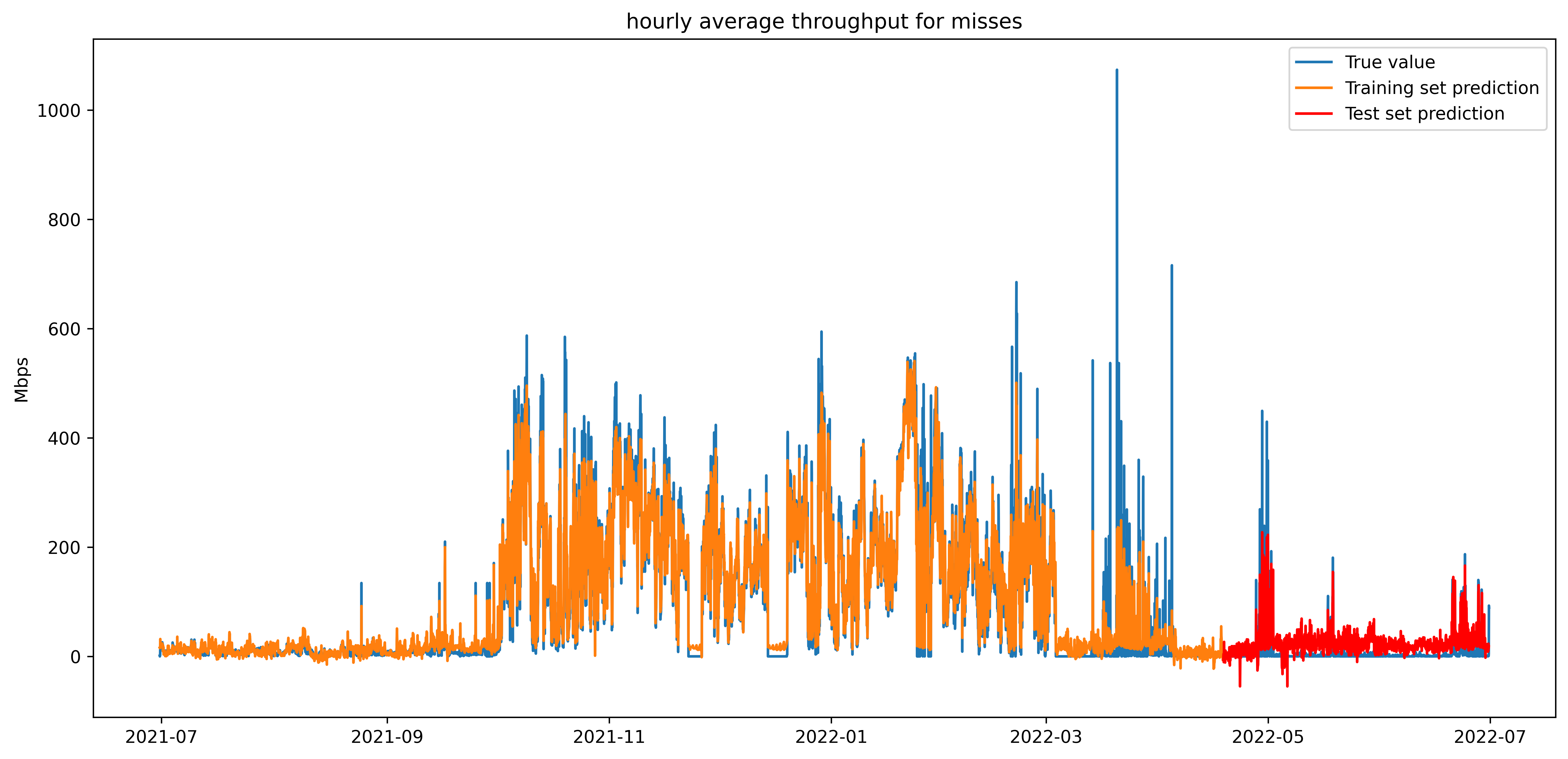}
    \label{fig:hourly_miss_tput}
}} 
\caption{Average throughput of wide-area file transfers}
\label{fig_miss_tput}
\end{figure}

\autoref{fig_miss_counts} and \autoref{fig_miss_size} show LSTM model
output for the number of cache misses and their associated data volumes.
In these cases, we see the predictions on the hourly time series are
closer to the actual values in the last period of time than the
predictions with daily time series,
even though the hourly time series often shows stronger spikes.

\autoref{fig_miss_tput} shows the LSTM model performance with daily and
hourly average throughput values.  From Figure~\ref{fig:daily_miss_tput}, we
see that during the middle of the observation period (10/21 -- 2/22),
the wide-area network traffic throughputs are quite high  because
the network transfers are dominated by relatively large files that are
typically better able to utilize the network capacity.  In the other
time periods, the average throughput is relatively low due to small
files being transferred.

In the hourly model from Figure~\ref{fig:hourly_miss_tput}, there are
significant number of spikes during late March and early May.  
Examining Figures~\ref{fig:hourly_miss_counts} and \ref{fig:hourly_miss_size}, we see that these spikes
occur during time periods with very few cache misses, i.e., very few wide-area
file transfers.
We are interested in exploring these spikes further in the future.
For modeling network performance, it is not necessary for us to capture
such spikes precisely.  We could use the moving average method to smooth out the spike to obtain a performance model for the general trends.

\begin{figure}
\centerline{%
\subfloat[Network throughput with 24-hour moving average]{%
    \includegraphics[width=0.8\linewidth]{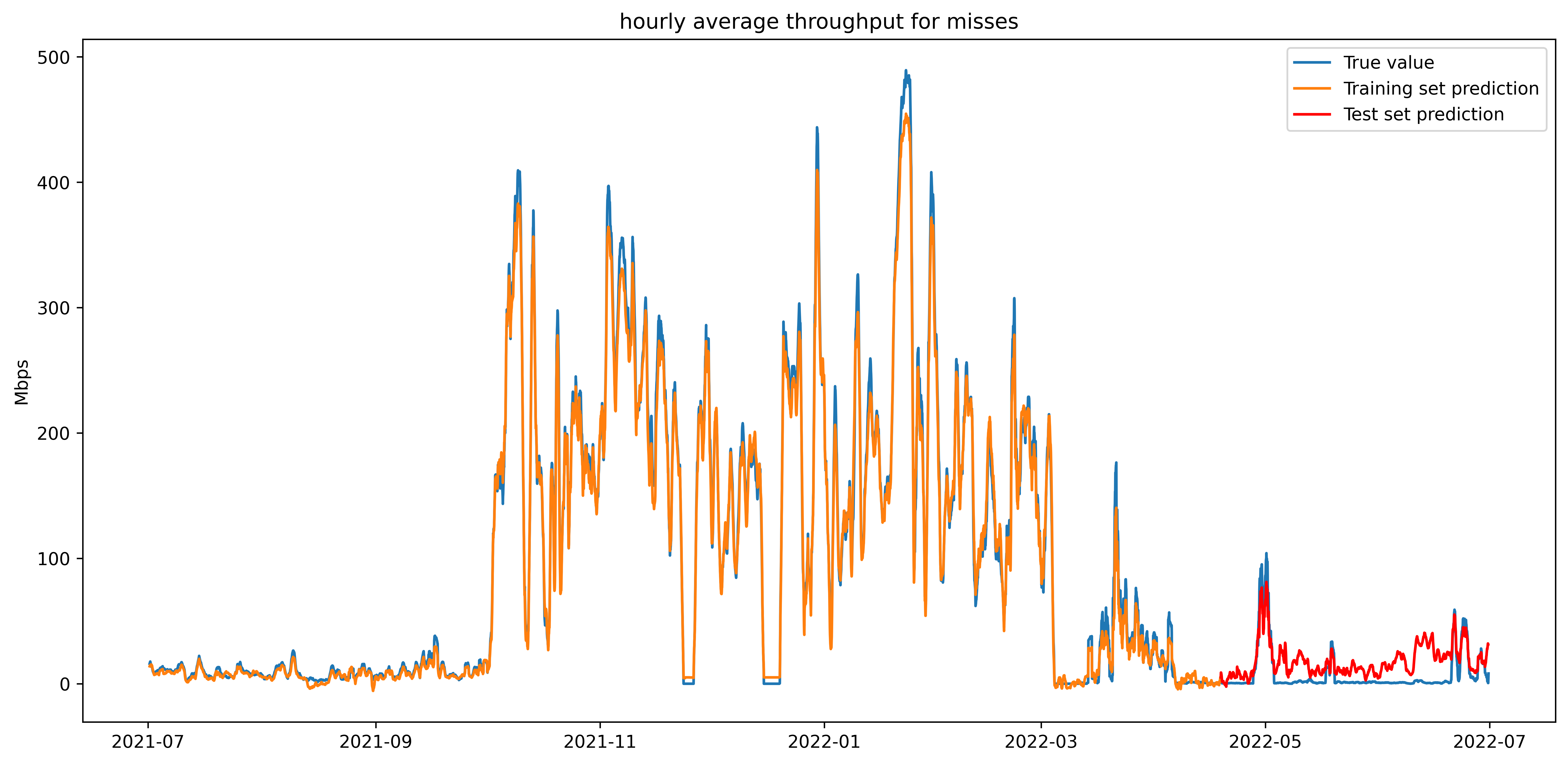}
    \label{fig:hourly_tput_1dma}
}}
\centerline{%
\subfloat[Network throughput with 168-hour moving average]{%
    \includegraphics[width=0.8\linewidth]{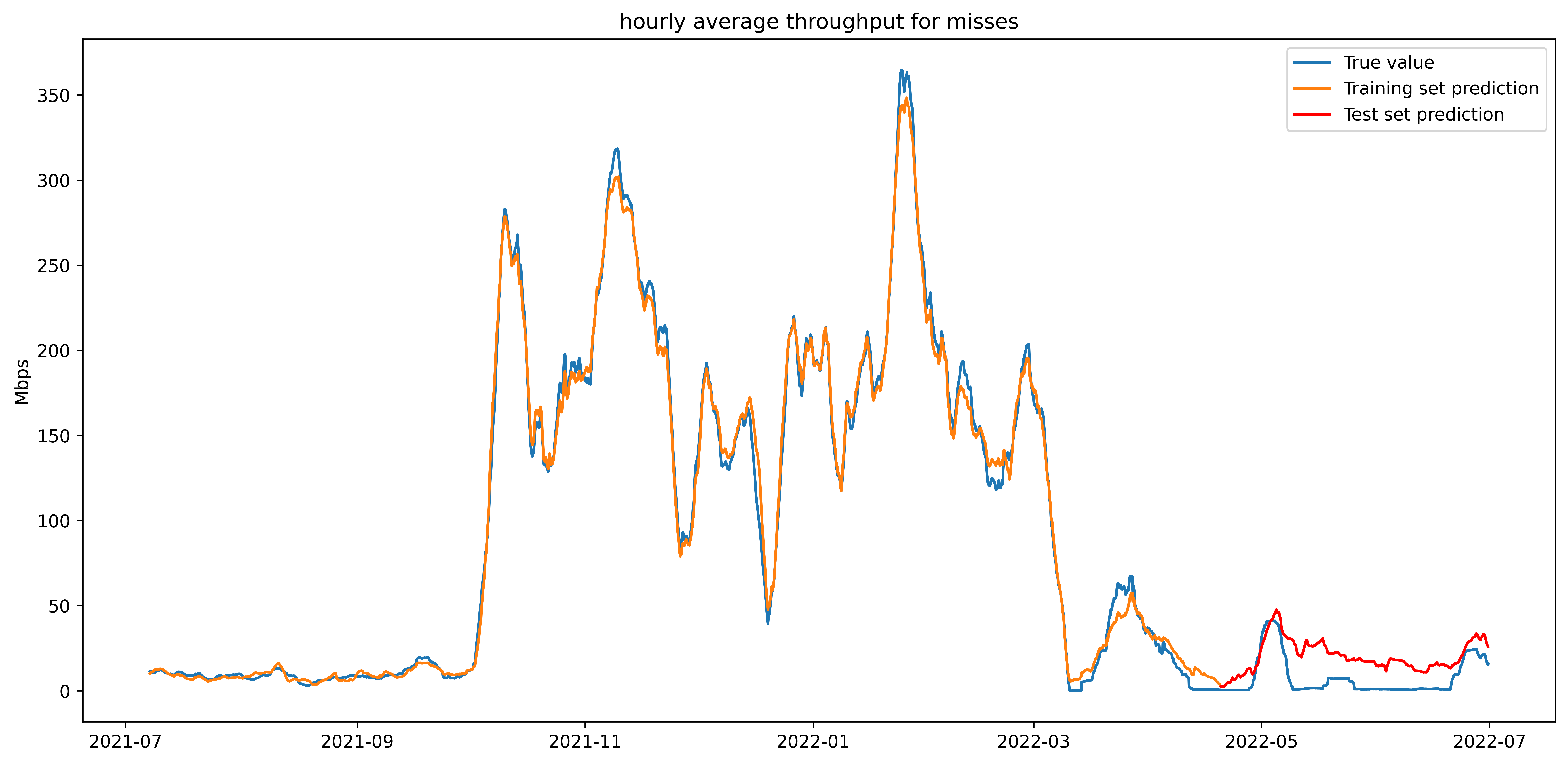}
    \label{fig:hourly_tput_7dma}
}} 
\caption{Modeling hourly average wide-area throughput on smoothed time series.}
\label{fig:tput_ma}
\end{figure}

\autoref{fig:tput_ma} shows two different versions of
Figure~\ref{fig:hourly_miss_tput} with two different moving-averaged hourly
throughput measured by the cache misses.  We clearly see that LSTM results match the
moving-averages much better than the original time series shown in Figure \ref{fig:hourly_miss_tput}.
The testing error (RMSE) of the LSTM predictions on the 24-hour
moving-averages is 15.05 (i.e., model in Figure~\ref{fig:hourly_tput_1dma})
and corresponding error on the 168-hour moving average is 14.56.
Both of these errors are less than 22.79 on the original hourly
throughput time series.
Even though the 24-hour moving averages look like the daily throughput time series shown in Figure~\ref{fig:daily_miss_tput}, the LSTM predictions matches the 24-hour moving averages much better based on visual inspection.
The RMSE of 15.05 (Figure~\ref{fig:hourly_tput_1dma}) is noticeably smaller than 23.49 (Figure~\ref{fig:daily_miss_tput}).
For anticipating future network performance, the LSTM model based on the moving averages is likely to work better.

%% file: xcache-icnc23.bbl
\begin{thebibliography}{10}
\providecommand{\url}[1]{#1}
\csname url@samestyle\endcsname
\providecommand{\newblock}{\relax}
\providecommand{\bibinfo}[2]{#2}
\providecommand{\BIBentrySTDinterwordspacing}{\spaceskip=0pt\relax}
\providecommand{\BIBentryALTinterwordstretchfactor}{4}
\providecommand{\BIBentryALTinterwordspacing}{\spaceskip=\fontdimen2\font plus
\BIBentryALTinterwordstretchfactor\fontdimen3\font minus
  \fontdimen4\font\relax}
\providecommand{\BIBforeignlanguage}[2]{{%
\expandafter\ifx\csname l@#1\endcsname\relax
\typeout{** WARNING: IEEEtran.bst: No hyphenation pattern has been}%
\typeout{** loaded for the language `#1'. Using the pattern for}%
\typeout{** the default language instead.}%
\else
\language=\csname l@#1\endcsname
\fi
#2}}
\providecommand{\BIBdecl}{\relax}
\BIBdecl

\bibitem{esnetHepReq}
\BIBentryALTinterwordspacing
B.~Brown, E.~Dart, G.~Rai, L.~Rotman, and J.~Zurawski, ``Nuclear physics
  network requirements review report,'' Energy Sciences Network, University of
  California, Publication Management System Report LBNL-2001281, 2020.
  [Online]. Available: \url{http://www.es.net/assets/Uploads/20200505-NP.pdf}
\BIBentrySTDinterwordspacing

\bibitem{datalakes}
\BIBentryALTinterwordspacing
X.~Espinal, S.~Jezequel, M.~Schulz, A.~Sciaba, I.~Vukotic, and F.~Wuerthwein,
  ``The quest to solve the hl-lhc data access puzzle,'' \emph{{EPJ} Web of
  Conferences}, vol. 245, p. 04027, 2020. [Online]. Available:
  \url{http://doi.org/10.1051/epjconf/202024504027}
\BIBentrySTDinterwordspacing

\bibitem{xrootdcms}
L.~Bauerdick, D.~Benjamin, K.~Bloom, B.~Bockelman, D.~Bradley, S.~Dasu,
  M.~Ernst, R.~Gardner, A.~Hanushevsky, H.~Ito, D.~Lesny, P.~McGuigan,
  S.~McKee, O.~Rind, H.~Severini, I.~Sfiligoi, M.~Tadel, I.~Vukotic,
  S.~Williams, F.~Wurthwein, A.~Yagil, and W.~Yang, ``Using xrootd to federate
  regional storage,'' \emph{Journal of Physics: Conference Series}, vol. 396,
  no.~4, p. 042009, 2012.

\bibitem{Fajardo2020}
\BIBentryALTinterwordspacing
E.~Fajardo, D.~Weitzel, M.~Rynge, M.~Zvada, J.~Hicks, M.~Selmeci, B.~Lin,
  P.~Paschos, B.~Bockelman, A.~Hanushevsky, F.~W{\"{u}}rthwein, and
  I.~Sfiligoi, ``Creating a content delivery network for general science on the
  internet backbone using {XCaches},'' \emph{{EPJ} Web of Conferences}, vol.
  245, p. 04041, 2020. [Online]. Available:
  \url{http://doi.org/10.1051/epjconf/202024504041}
\BIBentrySTDinterwordspacing

\bibitem{Kosar:2011:stork}
T.~Kosar, M.~Balman, E.~Yildirim, S.~Kulasekaran, and B.~Ross, ``Stork data
  scheduler: Mitigating the data bottleneck in e-science,'' \emph{Philosophical
  Transactions of the Royal Society A: Mathematical, Physical and Engineering
  Sciences}, vol. 369, no. 1949, pp. 3254--3267, 2011.

\bibitem{Tierney:1999:cache}
B.~L. Tierney, J.~Lee, B.~Crowley, M.~Holding, J.~Hylton, and F.~L. Drake, ``A
  network-aware distributed storage cache for data intensive environments,'' in
  \emph{Proceedings. The Eighth International Symposium on High Performance
  Distributed Computing (Cat. No.99TH8469)}, 1999, pp. 185--193.

\bibitem{socalrepo2018}
E.~Fajardo, A.~Tadel, M.~Tadel, B.~Steer, T.~Martin, and F.~Wwrthwein, ``A
  federated xrootd cache,'' \emph{Journal of Physics: Conference Series}, vol.
  1085, p. 032025, 2018.

\bibitem{esnet2022}
``Energy sciences network,'' \url{https://www.es.net}, accessed: 2022-10-12.

\bibitem{copps2021}
E.~Copps, H.~Zhang, A.~Sim, K.~Wu, I.~Monga, C.~Guok, F.~Wurthwein, D.~Davila,
  and E.~Fajardo, ``Analyzing scientific data sharing patterns with in-network
  data caching,'' in \emph{4th ACM International Workshop on System and Network
  Telemetry and Analysis (SNTA 2021)}, ACM.\hskip 1em plus 0.5em minus
  0.4em\relax ACM, 2021.

\bibitem{han2022}
R.~Han, A.~Sim, K.~Wu, I.~Monga, C.~Guok, F.~Wurthwein, D.~Davila, J.~Balcas,
  and H.~Newman, ``Access trends of in-network cache for scientific data,'' in
  \emph{5th ACM International Workshop on System and Network Telemetry and
  Analysis (SNTA 2022)}, ACM.\hskip 1em plus 0.5em minus 0.4em\relax ACM, 2022.

\bibitem{greff2016lstm}
K.~Greff, R.~K. Srivastava, J.~Koutn{\'\i}k, B.~R. Steunebrink, and
  J.~Schmidhuber, ``Lstm: A search space odyssey,'' \emph{IEEE transactions on
  neural networks and learning systems}, vol.~28, no.~10, pp. 2222--2232, 2016.

\bibitem{sherstinsky2020fundamentals}
A.~Sherstinsky, ``Fundamentals of recurrent neural network (rnn) and long
  short-term memory (lstm) network,'' \emph{Physica D: Nonlinear Phenomena},
  vol. 404, p. 132306, 2020.

\bibitem{xrootd2005}
A.~Dorigo, P.~Elmer, F.~Furano, and A.~Hanushevsky, ``Xrootd - a highly
  scalable architecture for data access,'' \emph{WSEAS Transactions on
  Computers}, vol.~4, no.~4, pp. 348--353, 2005.

\bibitem{miniaod2019}
\BIBentryALTinterwordspacing
A.~Rizzi, G.~Petrucciani, and M.~Peruzzi, ``A further reduction in cms event
  data for analysis: the nanoaod format,'' \emph{EPJ Web Conf.}, vol. 214, p.
  06021, 2019. [Online]. Available:
  \url{http://doi.org/10.1051/epjconf/201921406021}
\BIBentrySTDinterwordspacing

\bibitem{cmsoffline}
{CMS Collaboration}, ``The cms offline workbook,''
  \url{https://twiki.cern.ch/twiki/bin/view/CMSPublic/WorkBook}, accessed: Oct.
  1st, 2022.

\bibitem{stashcache}
D.~Weitzel, M.~Zvada, I.~Vukotic, R.~Gardner, B.~Bockelman, M.~Rynge,
  E.~Hernandez, B.~Lin, and M.~Selmeci, ``Stashcache: A distributed caching
  federation for the open science grid,'' in \emph{PEARC '19: Proceedings of
  the Practice and Experience in Advanced Research Computing on Rise of the
  Machines (learning)}, 07 2019, pp. 1--7.

\bibitem{xcache2014}
L.~Bauerdick, K.~Bloom, B.~Bockelman, D.~Bradley, S.~Dasu, J.~Dost,
  I.~Sfiligoi, A.~Tadel, M.~Tadel, F.~Wuerthwein, A.~Yafil, and the
  CMS~collaboration, ``Xrootd, disk-based, caching proxy for optimization of
  data access, data placement and data replication,'' \emph{Journal of Physics:
  Conference Series}, vol. 513, no.~4, 2014.

\bibitem{Malik:2005:bypass}
T.~Malik, R.~Burns, and A.~Chaudhary, ``Bypass caching: making scientific
  databases good network citizens,'' in \emph{21st International Conference on
  Data Engineering (ICDE'05)}, 2005, pp. 94--105.

\end{thebibliography}
